\documentstyle[aps,prd]{revtex}

\newcommand{\bq}{\begin{equation}}
\newcommand{\eq}{\end{equation}}
\newcommand{\bqn}{\begin{eqnarray}}
\newcommand{\eqn}{\end{eqnarray}}
\newcommand{\nb}{\nonumber}
\newcommand{\lb}{\label} 

\begin{document} 
 
\title{Generalized Vaidya Solutions}  
 \author{ Anzhong Wang \thanks{e-mail address: wang@symbcomp.uerj.br}}
\address{Departamento de F\' {\i}sica Te\' orica,
Universidade do Estado do Rio de Janeiro, 
Rua S\~ ao Francisco Xavier $524$, Maracan\~ a,
Cep. $20550-013 $ Rio de Janeiro~--~RJ, Brazil }
\author{Yumei Wu \thanks{e-mail address: yumei@dmm.im.ufrj.br}}
\address{Instituto de Matem\'atica, Universidade Federal do Rio
de Janeiro, Caixa Postal $68530$, Cep. $21945-970$,
Rio de Janeiro~--~RJ, Brazil}

\maketitle

\begin{abstract}

A large family of solutions, representing, in general, spherically
symmetric Type II fluid, is presented, which
includes most of the known solutions to the Einstein field equations,
such as, the monopole-de Sitter-charged Vaidya ones.  

\end{abstract}

\vspace{.4cm}


 \vspace{1.cm}

In 1951, Vaidya \cite{V1951} found a solution that represents an
imploding (exploding) null dust fluid with spherical symmetry. Since
then, the solution has been intensively studied in gravitational
collapse \cite{J1992}.  In particular, Papapetrou \cite{P1984} first
showed that this solution can give rise to the formation of naked
singularities, and thus provides one of the earlier counterexamples to
the cosmic censorship conjecture \cite{P1969}. Later, the solution was
generalized to the charged case \cite{LSM1965}. The charged Vaidya
solution soon attracted lot of attention and has been studied in
various situations. For example, Sullivan and Israel \cite{SI1980} used
it to study the thermodynamics of black holes, and Kaminaga \cite{K1990}
used it as a classical model for the geometry of evaporating charged
black holes, while Lake and Zannias \cite{LZ1991} studied the
self-similar case and found that, similar to the uncharged case, naked
singularities can be also formed from gravitational collapse.  Quite
recently, Husian \cite{H1996} further generalized the Vaidya solution to
a null fluid with a particular equation of state. 
Husian's solutions have been lately used
as the formation of black holes with short hair \cite{BH1997}.

In this Letter we shall generalize the Vaidya solution to a more general
case, which include most of the known solutions to the Einstein field
equations, such as, the monopole-de Sitter-charged Vaidya solutions, and
the Husian solutions. The generalization comes from the observation that
the energy-momentum tensor (EMT) is linear in terms of the mass
function. As a result, the linear superposition of particular
solutions is also a solution of the Einstein field equations. To show
this, let us begin with the general spherically symmetric line element
\cite{BI1991}
\bqn
\lb{eq1}
ds^{2} &=& - e^{2\psi(v, r)} \left[1 - \frac{2m(v, r)}{r}\right]dv^{2} + 
2 \epsilon e^{\psi(v, r)}dvdr\nb\\
& & + r^{2} \left(d\theta^{2} + 
\sin^{2}\theta d\varphi^{2}\right),\;\;(\epsilon = \pm 1),
\eqn
where $m(v, r)$ is usually called the mass function, and related to the
gravitational energy within a given radius $r$ \cite{LZ1991,PI1990}.
When $\epsilon = + 1$, the null coordinate $v$ represents the Eddington
advanced time, in which $r$ is decreasing towards the future  along a
ray $v = Const.$ (ingoing), while when $\epsilon = - 1$, it represents
the Eddington retarded time, in which $r$ is increasing towards the
future  along a ray $v = Const.$ (outgoing).

In the
following, we shall consider the particular case where $\psi(v, r) = 0$.
Then, the non-vanishing components of the Einstein tensor are given by,
\bq
\lb{eq2}
G^{0}_{0} = G^{1}_{1} = - \frac{2 m'(v, r)}{r^{2}}, \;\;
G^{1}_{0} = \frac{2 \dot{m}(v, r)}{r^{2}},\;\;
G^{2}_{2} = G^{3}_{3} = - \frac{ m''(v, r)}{r},
\eq
where $\{x^{\mu}\} = \{v, r, \theta, \varphi\},\; (\mu = 0, 1, 2, 3)$,
and 
$$
\dot{m}(v, r) \equiv \frac{\partial m(v, r)}{\partial v}, \;\;\;\;\;\;
m'(v, r) \equiv \frac{\partial m(v, r)}{\partial r}.
$$ 
Combining Eq.(\ref{eq2}) with the Einstein field equations $G_{\mu\nu} =
\kappa T_{\mu\nu}$, we find that the corresponding EMT can be written in
the form \cite{H1996}
\bq
\lb{eq3}
T_{\mu\nu} = T^{(n)}_{\mu\nu} + T^{(m)}_{\mu\nu},
\eq
where
\bqn
\lb{eq4}
T^{(n)}_{\mu\nu} &=& \mu l_{\mu}l_{\nu},\nb\\
T^{(m)}_{\mu\nu} &=& (\rho + P) \left(l_{\mu}n_{\nu} + 
l_{\nu}n_{\mu}\right) + P g_{\mu\nu},
\eqn
and 
\bq
\lb{eq5}
\mu = \frac{2 \epsilon \dot{m}(v, r)}{\kappa r^{2}},\;\;
\rho = \frac{ 2 m'(v, r)}{\kappa r^{2}},\;\;
P = - \frac{ m''(v, r)}{\kappa r},
\eq
with $l_{\mu}$ and $n_{\mu}$ being two null vectors,
\bqn
\lb{eq6}
l_{\mu} &=& \delta^{0}_{\mu},\;\;\; 
n_{\mu} = \frac{1}{2}\left[1 - \frac{2m(v, r)}{r}\right]
\delta^{0}_{\mu} - \epsilon \delta^{1}_{\mu},\nb\\
l_{\lambda}l^{\lambda} &=& n_{\lambda}n^{\lambda} = 0, \;\;
l_{\lambda}n^{\lambda} = - 1.
\eqn
The part of the EMT, $T^{(n)}_{\mu\nu}$, can be considered as the
component of the matter field that moves along the null hypersurfaces $v
= Const.$ In particular, when $\rho = P = 0$, the solutions reduce to
the Vaidya solution with $m = m(v)$. Therefore, for the general case we
consider the EMT of Eq.(\ref{eq3}) as a generalization of the Vaidya
solution.

Projecting the EMT of Eq.(\ref{eq3}) to the orthonormal basis, defined by
the four vectors,  
\bq
\lb{eq7}
E_{(0)}^{\mu} = \frac{l_{\mu} + n_{\mu}}{\sqrt{2}},\;\;
E_{(1)}^{\mu} = \frac{l_{\mu} - n_{\mu}}{\sqrt{2}},\;\;
E_{(2)}^{\mu} = \frac{1}{r}\delta^{\mu}_{2},\;\;
E_{(3)}^{\mu} = \frac{1}{r\sin\theta}\delta^{\mu}_{3},
\eq
we find that
\bq
\lb{eq8}
\left[T_{(a)(b)}\right] = \left[
\begin{array}{lccl}
\frac{\mu}{2} + \rho& \frac{\mu}{2}& 0 & 0\\
\frac{\mu}{2} & \frac{\mu}{2} - \rho & 0 & 0\\
0 & 0 & P & 0\\
0 & 0 & 0& P\\
\end{array} \right],
\eq
which in general belongs to the Type II fluids defined in \cite{HE1973}.
The null vector $\l^{\mu}$ is a double null eigenvector of the EMT. For
this type of fluids, the energy conditions  are the following
\cite{HE1973}:

 a) {\em The weak and strong energy conditions}:
\bq
\lb{eq9}
\mu \ge 0, \;\;\; \rho \ge 0, \;\;\; P \ge 0,\; (\mu \not= 0).
\eq

b) {\em The dominant energy condition}:
\bq
\lb{eq10}
\mu \ge 0, \;\;\; \rho \ge P \ge 0,\; (\mu \not= 0).
\eq
Clearly, by properly choosing the mass function $m(v, r)$, these
conditions can be satisfied. In particular, when $ m = m(v)$, as we
mentioned previously, the solutions reduce to the Vaidya solution, and
the energy conditions (weak, strong, and dominant) all reduce to $\mu
\ge 0$, while when $m = m(r)$, we have $\mu = 0$, and the matter field
degenerates to type I fluid \cite{HE1973}. In the latter case, the
energy conditions become:

c) {\em The weak energy condition}:
\bq
\lb{eq10a}
\rho \ge 0, \;\;\; P + \rho \ge 0,  \;\;( \mu = 0).
\eq

d) {\em The strong energy condition}:
\bq
\lb{eq10b}
\rho + P \ge 0, \;\;\; P \ge 0,  \;\;( \mu = 0).
\eq

e) {\em The dominant energy condition}:
\bq
\lb{eq10c}
\rho \ge 0, \;\;\; - \rho \le P \le \rho  \;\;( \mu = 0).
\eq

Without loss of generality, we expand $m(v, r)$ in the powers of $r$,
\bq
\lb{eq11}
m(v, r) = \sum^{+ \infty}_{n = - \infty} a_{n}(v) r^{n},
\eq
where $a_{n}(v)$ are arbitrary functions of $v$ only.  Note that the sum of the
above expression should be understood as an integral, when the
``spectrum'' index $n$ is continuous. Substituting it into
Eq.(\ref{eq5}), we find
\bqn
\lb{eq12}
\mu &=& \frac{2\epsilon}{\kappa}\sum^{+ \infty}_{n = - \infty}
{\dot{a}_{n}(v)r^{n - 2}},\;\;\;
\rho = \frac{2}{\kappa}\sum^{+ \infty}_{n = - \infty}
{n{a}_{n}(v)r^{n - 3}},\nb\\
P &=& - \frac{1}{\kappa}\sum^{+ \infty}_{n = - \infty}
{n(n- 1)a_{n}(v)r^{n - 3}}.
\eqn
The above solutions include most of the known solutions of the Einstein
field equations with spherical symmetry:

i) {\bf The monopole solution} \cite{BV1989}: If we choose the functions
$a_{n}(v)$ such that
\bq
\lb{eq13}
a_{n}(v) = \left\{
\begin{array}{ll}
\frac{a}{2},& n = 1,\\
0, & n \not= 1,
\end{array}\right.
\eq
where $a$ is an arbitrary constant, then we find
\bqn
\lb{eq14}
m(v, r) & = & \frac{a r}{2},\nb\\
\rho &=& \frac{a}{\kappa r^{2}},\;\;\;\; \mu = P = 0.
\eqn
Clearly, in this case the matter field is type I and satisfies all the
three energy conditions (\ref{eq10a}) - (\ref{eq10b}) as long as $a >
0$.  The corresponding solution can be identified as representing  the
gravitational field of a monopole \cite{BV1989} (see also \cite{L1979}).

ii) {\bf The de Sitter and Anti-de Sitter solutions}: If the functions
$a_{n}(v)$ are chosen such that
\bq
\lb{eq15}
a_{n}(v) = \left\{
\begin{array}{ll}
\frac{\Lambda}{6},& n = 3,\\
0, & n \not= 3,
\end{array}\right.
\eq
we find that
\bqn
\lb{eq16}
m(v, r) & = & \frac{\Lambda}{6} r^{3},\nb\\
\rho &=& - P = \frac{\Lambda}{\kappa},\;\;\;\; \mu =  0,
\eqn
and that
\bq
\lb{eq17}
T_{\mu\nu} = - \frac{\Lambda}{\kappa} g_{\mu\nu}.
\eq
This corresponds to the de Sitter solutions for $\Lambda > 0$, and to
Anti-de Sitter solution for $\Lambda < 0$, where $\Lambda$ is the
cosmological constant.

iii) {\bf The charged Vaidya solution}: To obtain the charged Vaidya solution,
we shall choose the functions $a_{n}(v)$ such that,
\bq
\lb{eq18}
a_{n}(v) = \left\{
\begin{array}{ll}
f(v), & n = 0,\\
- \frac{q^{2}(v)}{2},& n = - 1,\\
0, & n \not= 0, - 1,
\end{array}\right.
\eq
where the two arbitrary functions $f(v)$ and $q(v)$ represent,
respectively, the mass and electric charge at the advanced (retarded)
time $v$. Inserting the above expression into Eq.(\ref{eq12}), we find
that
\bqn
\lb{eq19}
m(v, r) & = & f(v) - \frac{q^{2}(v)}{2 r},\nb\\
\mu &=& \frac{2\epsilon}{\kappa r^{3}}\left[r \dot{f}(v) - 
q(v)\dot{q}(v)\right],\nb\\
\rho &=& P = \frac{q^{2}(v)}{\kappa r^{4}}.
\eqn
This is the well-known charged Vaidya solution. $T^{(n)}_{\mu\nu}$
corresponds to the EMT of the Vaidya null fluid, and $T^{(m)}_{\mu\nu}$
to the electromagnetic field, $F_{\mu\nu}$, given by,
\bq
\lb{eq20}
F_{\mu\nu} = \frac{q(v)}{r^{2}}(\delta^{0}_{\mu}\delta^{1}_{\nu}
- \delta^{1}_{\mu}\delta^{0}_{\nu}). 
\eq
From Eq.(\ref{eq19}) we can see that the condition $\mu \ge 0$
gives the main restriction on the choice of the functions $f(v)$ and
$q(v)$. In particular, if $df/dq > 0$, we can see that there always
exists a critical radius $r_{c}$ such that when $ r < r_{c}$, we have
$\mu < 0$, where
\bq
\lb{eq21}
r_{c} = q(v)\frac{\dot{q}(v)}{\dot{f}(v)}.
\eq
Thus, in this case the energy conditions are always violated. However, a
closer investigation of the equation of motion for the massless charged
particles that consist of the charged null fluid showed that in this
case the hypersurface $r = r_{c}$ is actually a vanishing point
\cite{O1991}. In the imploding case ($\epsilon = + 1$), for example, due
to the repulsive Lorentz force, the 4-momenta of the particles vanish
exactly on $r = r_{c}$. Afterwards, the Lorentz force will push the
particles to move outwards. Therefore, in realistic situations the
particles cannot get into the region $r < r_{c}$, whereby the energy
conditions are preserved \cite{O1991}.

iv) {\bf The Husian solutions}: If we choose the functions $a_{n}(v)$ 
such that
\bq
\lb{eq22}
a_{n}(v) = \left\{
\begin{array}{ll}
f(v), & n = 0,\\
- \frac{g(v)}{2k - 1},& n =2k - 1\; (k \not= 1/2),\\
0, & n \not= 0, 2k - 1,
\end{array}\right.
\eq
where $f(v)$ and $g(v)$ are two arbitrary functions, and $k$ is a
constant, then we find that
\bqn
\lb{eq23}
m(v, r) & = & f(v) - \frac{g(v)}{(2k -1)r^{2k - 1}},\nb\\
\mu &=& \frac{2\epsilon}{\kappa r^{2}}\left[ \dot{f}(v) - 
\frac{\dot{g}(v)}{(2k - 1)r^{2k - 1}}\right],\nb\\
P &=& k \rho = \frac{2kg(v)}{\kappa r^{2k + 2}}.
\eqn
This is the solution first found by Husian by imposing the equation of state
$P = k \rho$ \cite{H1996}. When $k = 1$,
they reduce to the charged Vaidya solution. Similar to the latter case, now
the condition $\mu \ge 0 $ also gives the main restriction on the choice
of the functions $f(v)$ and $g(v)$, especially for the case where $df/dg > 0$. 
However, one may follow Ori \cite{O1991} to argue that the hypersurface
$$
r = r_{c} = \left[(2k - 1)^{-1}\frac{dg}{df}\right]^{\frac{1}{2k - 1}},
$$
is also a turning point, although we have not been able to show this
explicitly. But the following considerations indeed support this point
of view. Following \cite{BH1997}, we can cast $T^{(m)}_{\mu\nu}$ into
the form of a {\em generalized} electromagnetic field,
\bq
\lb{eq24}
T^{(m)}_{\mu\nu} = \frac{2}{\kappa}\left(F_{\mu\lambda}F_{\nu}^{\;\lambda}
- \frac{\alpha}{4} g_{\mu\nu} F_{\lambda\sigma}F^{\lambda\sigma}\right),
\eq
where $\alpha = 2/(1+k)$, and $F_{\mu\nu}$ can be considered as the
generalized electromagnetic field, given by,
\bq
\lb{eq25}
F_{\mu\nu} = \frac{[k(1+k)m'(v, r)]^{1/2}}{r}
(\delta^{0}_{\mu}\delta^{1}_{\nu}
- \delta^{1}_{\mu}\delta^{0}_{\nu}),
\eq
which satisfies the Maxwell field equations,
\bq
\lb{eq26}
F_{[\mu\nu;\lambda]} = 0,\;\;\;\; 
F_{\mu\nu;\lambda}g^{\nu\lambda} = J_{\mu},
\eq
with
\bqn
\lb{eq27}
J_{\mu} &=& J_{0}\delta^{0}_{\mu} + J_{1}\delta^{1}_{\mu},\nb\\
J_{0} &=& \frac{2\delta q^{k+1}(v)}{r^{3(k+1)}}
\left\{k\dot{q}(v) r^{2(k+1)}
\right. \nb\\
& & \left. + (1 - k)r q(v)[q^{2k}(v) - 2f(v)r^{2k - 1} + 
r^{2k}]\right\},\nb\\
J_{1} &=& - \frac{2\delta (1 - k) q^{k}(v)}{r^{k + 2 }},\nb\\
g(v) &=& \frac{(2k-1)q^{2k}(v)}{2}, ( k \not= 1/2),
\eqn
where $\delta \equiv [k(1+k)(2k-1)/2]^{1/2}$. When $f$ and $g$ are
constants, from Eq.(\ref{eq23}) we have $\mu = 0$. Then, the solutions
degenerate to type I solutions, and the energy conditions (\ref{eq10a})
- (\ref{eq10c}) become, respectively, $g \ge 0, \; k \ge - 1$ for the
weak energy conditon, $ g \ge 0,\; k \ge 0$ or $ g\le 0, \; k \le -1 $
for the strong energy condition, and $g \ge 0,\; - 1 \le k \le + 1$ for
the dominant energy condition. Note that when $k > 1$, the ``supercharge"
$q$ has no contribution to the surface intergral at spatial infinity due
to the rapid fall off ($r^{- 2k}$) in the metric coefficients. Therefore,
it acts like short hair \cite{BH1997}. However, the existence of this
kind of hairs can be limited by the dominant energy condition.

Note that the functions $\mu, \; \rho$ and $P$ are linear in terms of
the derivatives of $m(v, r)$. Thus, the linear superposition of Cases i)
- iv) is also a solution to the Einstein field equations. In particular,
the combination,
\bq
\lb{eq28}
m(v, r) = \frac{a r}{2} + \frac{\Lambda}{6}r^{3} + f(v) 
- \frac{q^{2}(v)}{2 r},
\eq
would represent the monopole-de Sitter-charged Vaidya solutions. 
Obviously, by properly choosing the functions $a_{n}(v)$, one can obtain
as many solutions as wanted. The physical and mathematical properties of
these solutions will be studied somewhere else.


\end{document}